# Orbital Angular Momentum preserving guided mode in helically twisted hollow core photonic crystal fiber at Dirac point

Rik Chattopadhyay and Shyamal K. Bhadra*

School of Chemical Sciences, Indian Association for the cultivation of science, 2A & 2B Raja S. C. Mullick Road, Kolkata 700032, India
email: rcskb@iacs.res.in

**Abstract**: We report trapping and propagation of photonic Dirac mode in a helically twisted hollow core photonic crystal fiber (HC-PCF) where the trapped light in the hollow (air) defect can preserve the orbital angular momentum (OAM). We show that a photonic Dirac point can emerge even in a twisted system for a suitable choice of curvilinear coordinate and the related waveguide defect modes defined in the new basis can preserve the associated OAM during axial translation. The effect of twist rate, defect geometry and crystal dimension on the propagation of OAM carrying trapped Dirac modes is critically analyzed. The results derived by FEM simulation are verified with an analytical theory based on dynamics of Bloch modes in twisted photonic crystals which are in good agreement. The proposed HC-PCF can play an important role in exciting and guiding of OAM carrying modes that help particle trapping and quantum communication.

## 1. Introduction:

Orbital angular momentum (OAM) carrying light has found immense application in fields like particle trapping, quantum information storing, high bandwidth communication channel and quantum communication [1, 2]. Since the first proposal of exciting light that can carry certain OAM [3], during last three decades some reports suggested methods of exciting light that carries OAM which is useful for specific application. However, there is not much reported results related to confinement and propagation of such OAM carrying light over a long distance [4-8]. Though some of these specially designed waveguides can transmit OAM carrying modes over a few meters [5] but the excitation of such modes in the system needs external arrangements and it requires specially designed couplers to transmit the signal from the waveguide to the peripherals.

Recently it has been shown that when a solid core photonic crystal fiber (PCF) is twisted helically along the length the polarization degeneracy of the core modes gets lifted [8-10] and the photonic crystal (PC) cladding can support Bloch modes carrying certain OAM [11, 12]. As stated in [13, 14] the hexagonal cladding in normal PCF supports fundamental space filling mode (FSM) whose axial Poynting vector directs towards the fiber axis. When the PCF is slightly twisted the FSM is forced to follow a helical path around the axis and it picks up a transverse momentum. The discrete OAM order is then determined by the PC radius and twist rate [15]. It has also been shown that in such twisted PCF the core guided mode can resonantly couple with selected OAM carrying Bloch modes and as a result it suffers a series of dips in the output spectrum [13, 15]. It has also been shown that OAM can be preserved in the core mode if the core of the PCF is designed as a three bladed Y-shaped structure [14].

Our idea is to avoid such special design of PCF cores so that any standard fiber optic peripherals can be used while designing a system. This is achieved when we can couple the OAM carrying Bloch modes of a twisted PC in the defect core of a PCF. In previous reports it has been demonstrated that in some hexagonal PCs with typical values of pitch, Λ (centre to centre distance of two adjacent air holes) and air hole diameter, d, a conical Dirac point appears in the photonic band structures and PC modes at this Dirac point (DP) can be trapped in a centrally located air defect [16, 17]. The trapping of such PC modes in centrally located defect, deliberately embedded in the PC, can be tuned by changing the diameter or dielectric permittivity of the defect. We try to investigate the possibility of exciting PC modes in a helically twisted hollow core PCF (HC-PCF) cladding so that some topological phases can be associated with PC modes. Then by utilizing the Dirac point resonance (DR) between PC mode and defect mode we try to couple the phase information from PC mode to the defect mode. We expect that this phase information will be preserved by the propagating defect modes in the air-core of HC-PCF. Hence we can achieve the goal of exciting a guided mode carrying certain OAM in a HC-PCF through this process.

We propose a helically twisted HC-PCF with $Λ = 2.21\ μm$ and $d = 0.94Λ$ with a centrally located air defect of radius $r_D = 1.73Λ$ to excite and guide OAM modes with OAM order ±1. We have studied three different fiber with host glass as pure fused silica $ε_r = 2.1025$, $GeO_2$ $ε_r = 2.5281$ and $SF_6$ $ε_r = 3.24$ [16]. In the present case, the Bloch modes at Dirac frequency are forced to follow a helical path due to twist. Hence, the Bloch modes will rotate during propagation and the rotation frequency will depend on the twist rate [15]. Now the spiraling Bloch modes are trapped in the defect when a resonance occurs between the defect mode and Bloch mode at Dirac frequency. This in turn excites a guided mode at the central hollow core. The resonance will be maintained if the defect mode started twisting at same frequency as that of the Bloch mode. Only at such condition the defect mode will see a consistent Dirac point during propagation. Hence the defect mode will pick up an azimuthal momentum and in turn carry some OAM order. The Bloch modes rotating along the twist experiences a Dirac point resonant guiding mechanism. On the other hand, Bloch modes rotating against the twist cannot see the Dirac cone like degeneracy as the translation invariance of the PC is lifted. Therefore, these modes cannot resonate with a defect mode. Since the twist induces circular birefringence between the degenerate polarizing modes [8, 18], the defect modes in helically twisted HC-PCF show transmission peaks at different wavelengths. Therefore it becomes easy to separate the different OAM modes in the proposed fiber.

We use mode analysis technique in finite element method (FEM) based COMSOL multiphysics to calculate the photonic bandgap and bandgap dispersion of the PCs as considered in the present work. The dispersion and transmission loss of the defect modes at Dirac frequencies are also calculated using FEM. The transmission window for the OAM modes are also calculated using an analytical theory based on the behavior of Bloch modes in twisted crystal and the results are verified with the transmission loss calculated using FEM. The effects of twist rate, PC dimension, host glass permittivity and geometry of the defect are studied and a universal design methodology is proposed. We show that the transmission window of such guided OAM modes can be tuned by the external twist introduced in the HC-PCF.

## 2. Dirac point in twisted HC-PCF:

It has been already established that the photonic band structure shows a conical Dirac degeneracy for a suitable value of d and Λ [16, 17]. The pertinent question arises in case of twisted HC-PCF (as shown in the Fig.-1) whether the Bloch mode degeneracy of the PC will hold or not? It has been established earlier that when Z-invariance for two dimensional PC is lifted, the frequency degeneracy of Bloch modes also gets lifted [8–10]. Therefore, it is expected that the Dirac like degeneracy of the Bloch modes will be lifted in such twisted HC-PCF. But if we make a coordinate transformation from rectangular Cartesian system (X,Y,Z) to a helical system (U,V,W) then the Z-invariance can be restored. Equation 1 gives the coordinate transformation relation as follows,

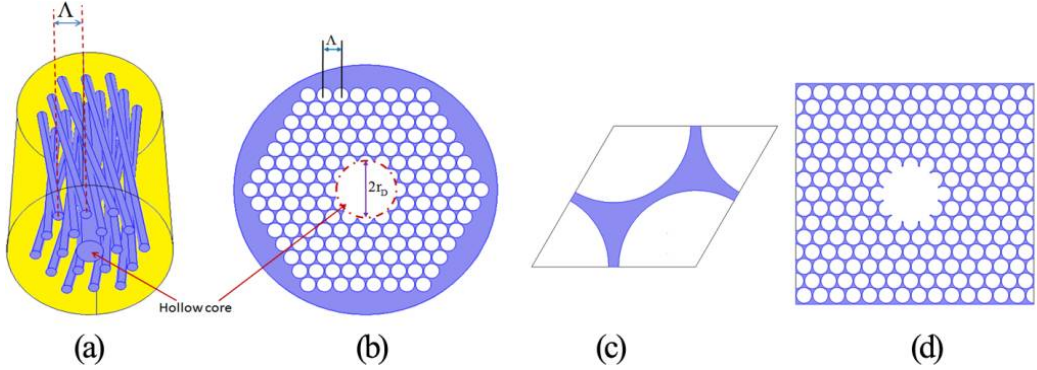

Figure 1. (a) Schematic of the helically twisted HC-PCF. The blue cylinders represent the air capillaries. The central large capillary serve as the hollow air-core. The yellow region indicates the base glass. (b) Cross-sectional view of the actual geometry considered in the study. The defect is indicated by the dashed red circle with radius $r_D = 1.73\Lambda$. (c) Unit cell of the PC. The white region is air and blue is base glass. (d) 14×14 supercell with central defect used for calculating the defect mode.

$$\begin{pmatrix} X \\ Y \\ Z \end{pmatrix} = \begin{pmatrix} \cos\theta & \sin\theta & 0 \\ -\sin\theta & \cos\theta & 0 \\ 0 & 0 & 1 \end{pmatrix} \begin{pmatrix} U \\ V \\ W \end{pmatrix} \quad (1)$$

where the angle $\theta = \alpha z$, $\alpha$ is the twist rate given in unit of rad/mm and z is the length of propagation from input plane. Close inspection of Equation 1 reveals that the Maxwell's equation in such curvilinear coordinate (U,V,W) will no longer remains orthogonal i.e. the variation of the electric field along Z axis cannot be replaced by a simple propagation phase $e^{i\beta z}$ where $\beta$ is the propagation constant.

For an untwisted crystal, the reciprocal vectors for a hexagonal PC can be written as [19],

$$\begin{pmatrix} b_1 \\ b_2 \end{pmatrix} = \frac{2\pi}{\Lambda} \begin{bmatrix} 1 & -\frac{1}{\sqrt{3}} \\ 0 & \frac{2}{\sqrt{3}} \end{bmatrix} \begin{pmatrix} \hat{\imath} \\ \hat{\jmath} \end{pmatrix} \quad (2)$$

where $\hat{\imath}$ and $\hat{\jmath}$ are the unit vectors along X-axis and Y-axis respectively. Now if we consider a helically twisted PC with twist rate $\alpha$ then the reciprocal lattice vectors can be written as using the transformation Equation 1,

$$\begin{pmatrix} g_1 \\ g_2 \end{pmatrix} = \frac{2\pi}{\Lambda} \begin{bmatrix} \frac{\cos(\frac{\pi}{6}+\alpha z)}{\cos(\frac{\pi}{6})} & -\frac{\sin(\frac{\pi}{6}+\alpha z)}{\cos(\frac{\pi}{6})} \\ \frac{\sin(\alpha z)}{\cos(\frac{\pi}{6}+2\alpha z)} & \frac{\cos(\alpha z)}{\cos(\frac{\pi}{6}+2\alpha z)} \end{bmatrix} \begin{pmatrix} \hat{\imath} \\ \hat{\jmath} \end{pmatrix} \quad (3)$$

Now if we consider that Bloch mode exists for such a twisted PC then the electric field of the Bloch modes can be written as [19],

$$E(\vec{r}) = \sum_{m,n} E_{m,n}(\vec{r}) e^{-i(\vec{k'}+m g_1+n g_2)\cdot\vec{r}} \quad (4)$$

Where $E_{m,n}(\vec{r})$ denotes the amplitude of the electric field at $\vec{r}$ for (m,n) order Bloch mode. $\vec{k'} = \frac{2\pi}{\lambda}\hat{k}$ in general denotes the propagation vector along z for operating wavelength $\lambda$. We have plotted the in-plane Bloch vector $[\vec{g}_1, \vec{g}_2]$ and phase $\phi$, of the m$^{th}$ order Bloch mode (m=n) with propagation distance Z. The plots are shown in Fig.-2.

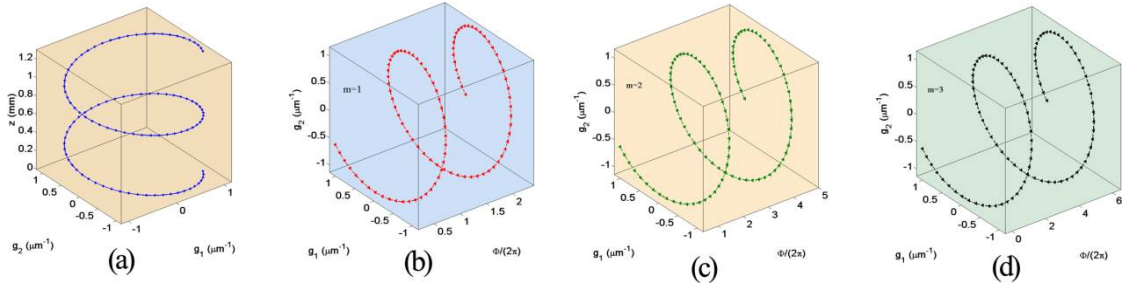

Figure 2. (a) Tip of the Bloch vector $[\vec{g}_1, \vec{g}_2]$ for each Z upto 2L where $L = 2\pi/\alpha$ mm is the twist pitch. Variation of the phase angle of the electric field with propagation length (b) for first higher order Bloch mode m=n=1 (c) for second higher order Bloch mode m=n=2 and (d) for third higher order Bloch mode m=n=3.

It is found that the Bloch vector traverses a helical path in the twisted PC. The phase associated with $m^{th}$ order Bloch mode varies from $0 \to 2\pi m$ over the transverse plane. Therefore, the Bloch modes in a twisted PC will carry an OAM. This indicates that though the PC appears anisotropic for the Bloch modes defined in a laboratory frame of reference, but the PC remains isotropic for the Bloch modes defined in a helical coordinate frame. Hence the DP will occur for such modes in the twisted PC. The only difference is that due to accumulation of some azimuthal phase the wave-vector of the Bloch modes will be raised by some amount, and the DR between the Bloch modes and defect modes will occur at different frequencies.

It is to be noted that the twist does not change the $d/\Lambda$ ratio of the PC nor the dielectric permittivity. Therefore the eigenfrequency of the Bloch modes i.e. photonic band structure will not differ in twisted and untwisted PC. The twist only forces the light to traverse a helical path to get confined in the glass between two adjacent air holes. This indicates that Bloch modes have to traverse a twisted path in order to conserve the transverse field distribution. Therefore the effective refractive index of the Bloch modes will increase. It has been shown earlier that in a coreless helically twisted PCF the index of the Bloch modes increases by the amount given by [15, 18],

$$n_{tw} = n_0\sqrt{(1 + \alpha^2 r^2)} \tag{5}$$

Where $n_{tw}$ is the effective index of the Bloch modes in twisted PC, $n_0$ is the effective index of Bloch mode in untwisted PC and $r$ is the distance from the fiber center. Assuming this relation still holds for a PC with 97% air filling fraction we estimated the change of effective refractive index of the Bloch modes for the PCs considered in the present study. This is demonstrated in Fig.-3. In this case we have considered a twisted HC-PCF with a circular centrally located defect that has a radius $r_D = 1.73\Lambda$ and the cladding PC consists of five rings with $\Lambda = 2.21\ \mu m$. Hence the maximum value of helix radius $r$ for the Bloch mode is $r_{max} < 7\Lambda \sim 15\ \mu m$ and the minimum helix radius will be $r_{min} > 2\Lambda \sim 5\ \mu m$. Fig.-3 shows the change of effective index of the Bloch modes that exist in the cladding with wavelength.

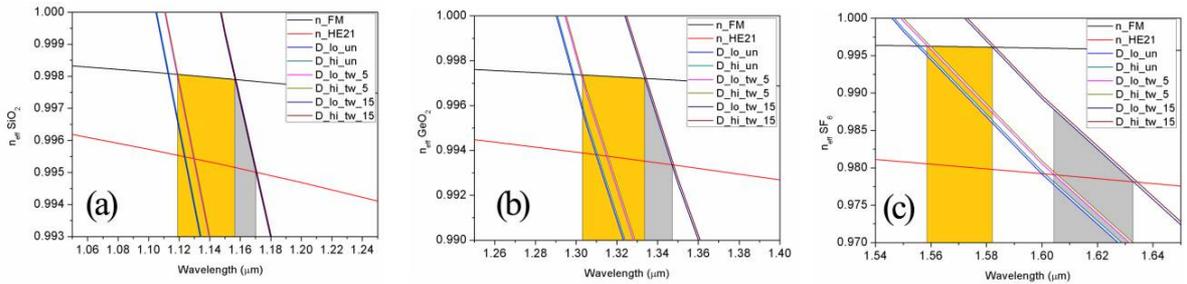

Figure 3. Dispersion of the fundamental (n_FM) and $HE_{21}$ (n_HE21) defect mode and PC modes at DP with and without twist calculated using FEM [13], [17] and Equation 5. (a) $SiO_2$ glass-air PC (b) $GeO_2$

glass-air PC and (c) $SF_6$ glass-air PC. D_lo_un and D_hi_un denote the 5$^{th}$ and 6$^{th}$ PC band at Dirac point for untwisted case respectively. D_lo_tw_5 and D_hi_tw_5 denote the PC bands at DP in twisted PC with helix radius r = 5 μm. D_lo_tw_15 and D_hi_tw_15 denote the PC bands at DP in twisted PC with helix radius r = 15 μm. The yellow region shows the DR region with fundamental defect mode and gray region denotes the DR region with $HE_{21}$ defect mode.

Since analytical solution of $E(\vec{r})$ is not possible in the twisted frame as we defined the electric field in a non-orthogonal curvilinear coordinate FEM is used to solve for the electric field. We consider the unit cell as shown in Figure 1(c) where the DP appears at point $K: (4\pi/3\Lambda, 0)$ of the irreducible Brillouin zone for all considered PCs [17]. Therefore, during FEM simulation we set the Floquet periodicity at the unit cell boundaries with periodic vector value $\vec{K} = 4\pi/3\Lambda \hat{\imath} + 0\hat{\jmath}$. We have calculated dispersion of the defect mode separately to identify DR in such twisted HC-PCF. The dispersion of the defect mode is calculated using the same method. The only difference is instead of taking a single unit cell we consider a 14×14 supercell with a defect inserted at the centre as shown in Figure 1(d). The dispersion of defect mode is also plotted in Fig.-3. The plot reveals that the DR occurs at a particular frequency for untwisted case. As soon as twist is introduced in the system, the effective mode index of the Bloch modes at DP takes a range of values depending on the value of r ranging from 5 μm to 15 μm (Equtaion 5). We also observe that when the dielectric permittivity contrast is high in PC, the dispersion of Dirac modes of the PC gets flattened and the DR region for different defect modes become well separated.

The results show that in twisted PC a DP may form if the Bloch modes follow the helical path induced by the twist. The coordinate transformation can restore the translation invariance and thus Bloch modes can be well defined in twisted PC. We also observe that the twist will increase the effective refractive index of the Bloch modes without altering the photonic band structure. As a result the DR will occur over a range of wavelength, rather than at particular frequency as in untwisted case. This study indicates that it is possible to trap photons from Bloch modes to a centrally located circular air defect in twisted HC-PCF and the transmission window at DR will widen.

**3. OAM modes in twisted HC-PCF:**

The nature of the trapped light in the hollow core of a twisted HC-PCF is now investigated using FEM. The non-orthogonality of the coordinate frame prevents to use the method of separation of variables to calculate the components of the electric field using a set of orthogonal equations. Hence no analytical solution of Maxwell's wave equation in helical coordinate system is possible. But it can be shown that if we change the material properties from an isotropic to anisotropic system using the transformation relation [7, 13, 20],

$$\epsilon' = \varepsilon T^{-1} \text{ and } \mu' = \mu T^{-1} \quad (6)$$

where

$$T^{-1} = \begin{bmatrix} 1 + \alpha^2 v^2 & -\alpha^2 uv & -\alpha v \\ -\alpha^2 uv & 1 + \alpha^2 u^2 & \alpha u \\ -\alpha v & \alpha u & 1 \end{bmatrix} \quad (7)$$

the wave equation can be reduced to a two dimensional problem and now it can be solved using FEM. We have used a perfectly matched layer (PML) to calculate the confinement loss of the core guided modes. OAM modes in such twisted PCF can be excited using the DR. We have already explained that in a twisted HC-PCF the Bloch mode follows a helical path and therefore carries certain orders of OAM. Now if we design the central defect in such a way that the $HE_{mn}$ for m ≥ 2 in the core shows a DR with any of the twisted Bloch modes then the core mode will carry an OAM by virtue of the topology.

We have simulated modal electric field for the twisted HC-PCF using FEM analysis where the permittivity tensor of the air capillaries were transformed using Equation 6 and 7 to capture their relative displacement in the glass. Fig.-3 depicts that the $HE_{21}$ mode in the defect shows DR from 1:14-1:17 μm in $SiO_2$ glass-air PC, 1:34-1:36 μm in $GeO_2$ glass-air PC and for $SF_6$ glass-air PC the DR region is in between 1:61-1:62 μm. The operating wavelengths are chosen from these resonance wavelength regions as shown in Fig.-3. The transverse field distribution plots for the defect modes with lowest confinement loss in these regions are shown in Fig.-4.

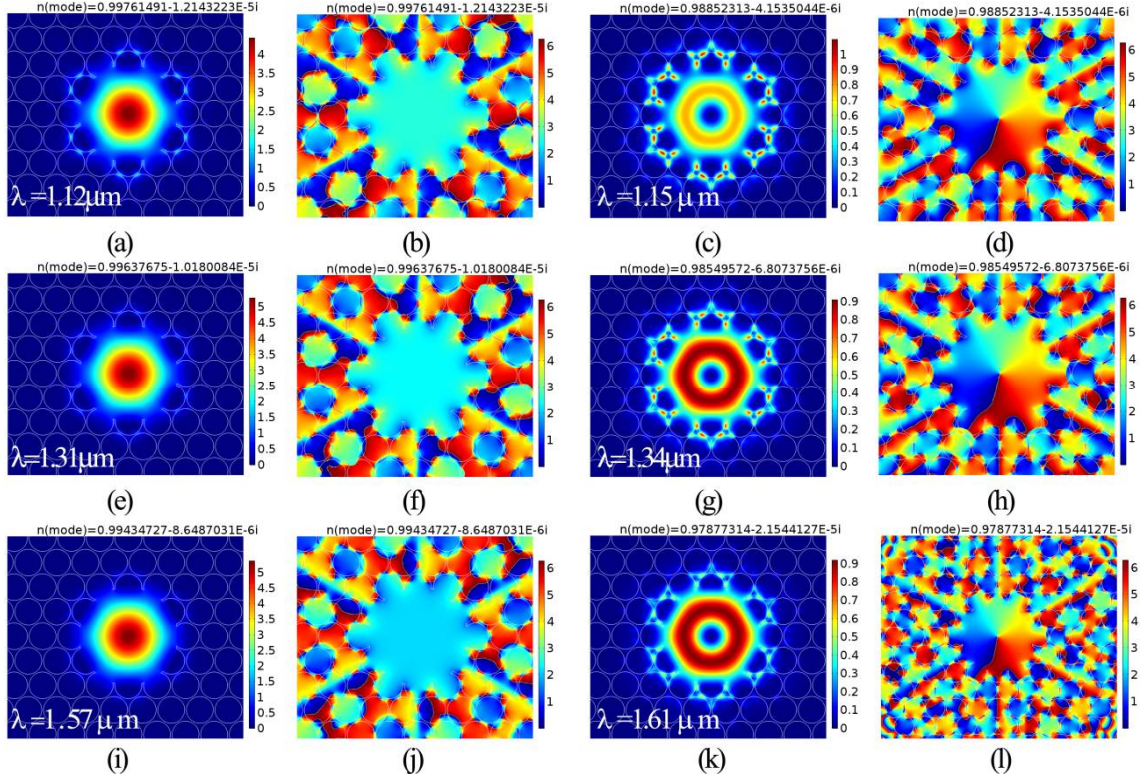

Figure 4. Power flow along z-axis of $V_{11}^+$ and $V_{21}^+$ mode along with the transverse phase distribution obtained by FEM analysis. $V_{11}^+$ power distribution for (a) $SiO_2$ glass-air PC (e) $GeO_2$ glass-air PC (i) $SF_6$ glass-air PC. $V_{11}^+$ phase distribution for (b) $SiO_2$ glass-air PC (f) $GeO_2$ glass-air PC (j) $SF_6$ glass-air PC. $V_{21}^+$ power distribution for (c) $SiO_2$ glass-air PC (g) $GeO_2$ glass-air PC (k) $SF_6$ glass-air PC. $V_{21}^+$ phase distribution for (d) $SiO_2$ glass-air PC (h) $GeO_2$ glass-air PC (l) $SF_6$ glass-air PC. The operating wavelengths are shown in the insets. The twist rate is set to $\alpha = 9.6 \ rad/mm$ i.e. in clockwise twist for all the cases.

Fig.-4 shows that the DR between PC modes and different defect modes are possible in the twisted HC-PCF. The phase distribution clearly reveals a phase singularity exists at the centre of the $V_{21}^+$ modes of the defect. We have ensured that the higher order defect modes carry OAM by plotting the x and y component of the transverse field (Fig.-5) as shown in Figure 4(k). Any OAM carrying mode can be written as [6],

$$V_{21}^+ = HE_{21}^{even} + jHE_{21}^{odd} \qquad (8)$$

where $j = \sqrt{-1}$. This also means, unlike a cylindrical vortex mode that has linearly polarized orthogonal components, the x and y components of $V_{21}^+$ will have intensity distribution of $HE_{21}$ mode (Fig.-5). The non-circularity of the component fields originate from the non-circular nature of the core [21, 22] and due to the thin glass extensions in the core due to inverse floral shape.

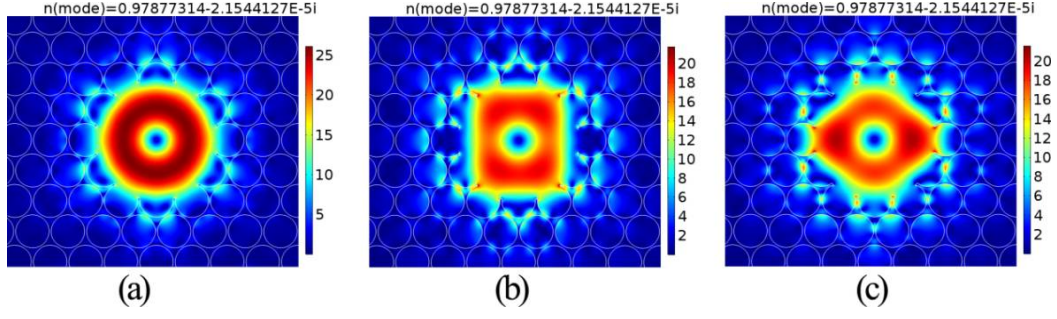

Figure 5. Transverse electric field intensity distribution of the mode shown in Fig. 4(k). (b) X-component of the field intensity (obtained by plotting abs(emw.Ex) in Comsol mode analysis) of the mode shown in (a). (c) y-component of the field intensity (obtained by plotting abs(emw.Ey) in Comsol mode analysis) of the mode shown in (a). The non-circularity of the component fields originate from the non-circular nature of the core and also due to the thin glass extensions in the core due to inverse floral shape.

The confinement loss is calculated using the relation [23],

$$\gamma = \frac{2\pi}{\lambda} Im\{n\} \qquad (9)$$

where $Im\{n\}$ denotes the imaginary part of the effective refractive index. Using Equation 8 we have calculated the transmittance of the fundamental and $HE_{21}$ mode in untwisted HC-PCF and OAM modes ($V_{11}^+$ and $V_{21}^+$) with order $l = 0, +1$ modes in twisted HC-PCF. The transmittance plots are shown in Fig.-6.

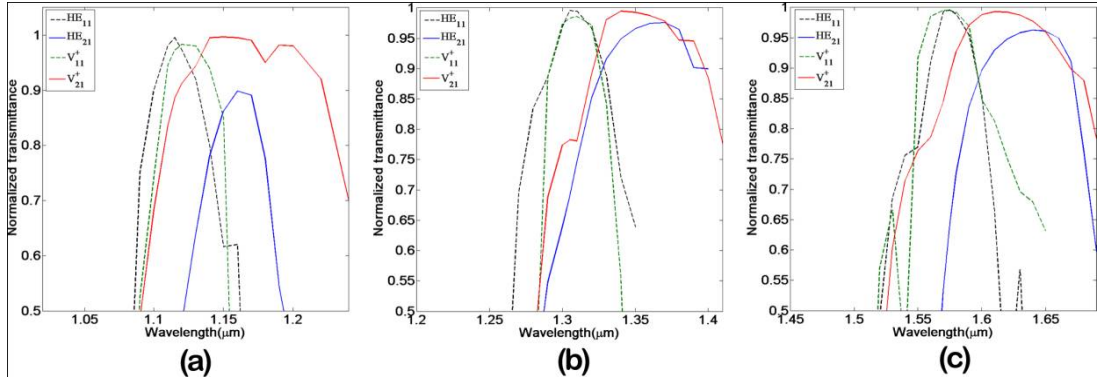

Figure 6. Transmittance plot of $HE_{11}$ and $HE_{21}$ defect modes for untwisted and $V_{11}^+$ and $V_{21}^+$ for twisted HC-PCF calculated using FEM (a) $SiO_2$ glass-air PC (b) $GeO_2$ glass-air PC (c) $SF_6$ glass-air PC.

Figure 6 shows that when the PCF is twisted the transmission windows of the $V_{11}^+$ and $V_{21}^+$ mode spread out. This can be correlated with the resonance plots as shown in Fig.-3. In untwisted case the DR occurs at a specific wavelength (Fig.-3). Thus we get a narrow transmission window bounded by the higher and lower Dirac band [16, 17]. On the other hand we observe that for twisted HC-PCF the Bloch modes participating in Dirac degeneracy show resonance with different defect modes for a band of wavelengths. Therefore, we expect that in twisted HC-PCF the maximum transmission will occur over this wider wavelength range (Fig.-6). The FWHM of transmission window increases for the twisted case for all considered HC-PCFs.

### 4. Dependency on geometrical parameters:
4.1. Core geometry
We have considered a centrally located circular defect placed in the PC to study the OAM excitation and guidance. Such a simple geometry is good for initial estimation and understanding the physical principles. In practice fabrication of such inverse floral shape (Fig.-1(b)) is bit difficult to achieve. Therefore we have extended our study for a more realistic core geometry which is found in regular HC-PCF, fabricated by stack and draw method. The cross-section of the geometry along with a mode field distribution for one of the considered twisted HC-PCF is shown in Fig.-7. It is evident (Fig.-7) that the nature of the trapped modes does not change with the change in shape of the hollow core. But the presence of a glass layer around the core increases the transmission loss of the trapped modes.

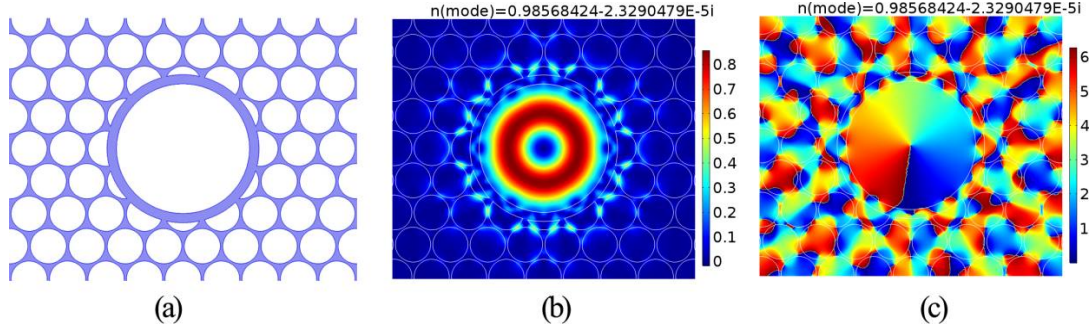

Figure 7. (a) Twisted HC-PCF with circular core in $GeO_2$ glass-air based PC. The core glass thickness is 0.24Λ. (b) Power distribution of $V_{21}^-$ mode at 1.36 μm. The twist rate α=9.6 rad/mm in anticlockwise direction. (c) Phase distribution of the mode shown in (b).

As a result the DR wavelength shows a red shift. This is a very trivial phenomenon. Thin glass around the air core acts as Fabrey-Perot cavity. Hence the intensity of the surface modes gets enhanced and thus the transmission loss of the trapped mode increases. The propagation loss of the OAM carrying mode at DR calculated by FEM for the three chosen HC-PCFs is given in Table 1.

Table 1. Propagation loss and DR wavelength for floral and circular core HC-PCF

|  | Floral core | | Circular core | |
| --- | --- | --- | --- | --- |
|  | DR Wavelength (μm) | Loss (dB/cm) | DR Wavelength (μm) | Loss (dB/cm) |
| $SiO_2$ glass-air | 1.15 | 1.97 | 1.16 | 5.66 |
| $GeO_2$ glass-air | 1.34 | 2.76 | 1.36 | 9.34 |
| $SF_6$ glass-air | 1.61 | 7.3 | 1.62 | 24.9 |

Therefore a special attention must be given in designing the core of the twisted HC-PCF. If we reduce the amount of glass around the core, we would achieve maximum propagation of such OAM carrying modes in a twisted HC-PCF.

4.2. PC size

PC size i.e. the number of air capillaries plays an important role in altering the confinement loss of the hollow core guided mode in conventional photonic bandgap HC-PCF. Therefore it is expected that if we increase the number of capillary layers in the PC cladding then the OAM modes will show lesser transmission loss. In this study we consider 5 layers of air capillaries in the cladding PC. We have increased the number of capillary layers from 5 to 7 in the PC. The transmission loss of the OAM modes in the twisted HC-PCF with 7 capillary layers at DR point are given in Table 2.

We observe that when the layer number is increased the confinement loss of the OAM carrying modes decreases and shows a little red shift. This can be explained based on results plotted in Figure 3. When the ring number is increased, the upper wavelength limit of the OAM pass band in twisted fiber will show a red shift. Therefore the maximum DR point will be shifted owing to large values of PC radius. But we must keep in mind about the structural integrity while increasing the number of air rings in PC. Twist may generate tensile stress that would change the shape of the air capillaries from circular to elliptical and will destroy the isotropic nature of the PC. Since we are considering a PC with almost 97% air filling fraction, so incorporation of 8 or more rings might lead to twist induced crack in the HC-PCF during fabrication. So an optimization is needed during fabrication.

Table 2. Propagation loss and DR wavelength for 5 and 7 layers of air capillaries in PC with floral core

|  | 5 layers | | 7 layers | |
| --- | --- | --- | --- | --- |
|  | DR Wavelength (μm) | Loss (dB/cm) | DR Wavelength (μm) | Loss (dB/cm) |
| $SiO_2$ glass-air | 1.15 | 1.97 | 1.17 | 0.41 |
| $GeO_2$ glass-air | 1.34 | 2.76 | 1.35 | 0.47 |
| $SF_6$ glass-air | 1.61 | 7.3 | 1.62 | 1.84 |

## 5. Conclusions:

We showed that OAM carrying modes with acceptable purity could be excited and transmitted through a suitably designed twisted HC-PCF. Our study exemplified that a new guidance mechanism can take place in a twisted HC-PCF where OAM carrying Bloch modes first get excited in the PC cladding. Satisfying the Dirac resonance condition such Bloch modes can be trapped in a centrally located air core. The process convincingly shows that Dirac cone like degeneracy of Bloch modes can be preserved in a twisted PC under rotating frame of reference. This in turn induces a topological phase in the PC modes. As a result this OAM information can be extracted and transmitted via a guiding channel. We have analyzed the effect of different fiber parameters on the transmission of OAM carrying modes of particular order. We provided a guideline of fabricating such HC-PCFs using standard fiber fabrication techniques. Our design is universal and compatible with standard fiber optic peripherals. This type of optical fiber will provide a tool for generating and transmitting OAM carrying lights, particle trapping and quantum communication.

**Acknowledgements**

The authors would like to acknowledge Prof. Shantanu Bhattacharya, Director, IACS for his support to carry out the research work. The authors are indebted to CSIR-ES scheme-21(2017)/15/EMR-II for funding.